# HANDWRITTEN TEXT IMAGE AUTHENTICATION USING BACK PROPAGATION

*A S N Chakravarthy[†]* , *Penmetsa V Krishna Raja ††* , Prof. *P S Avadhani[†††]*

[†] Associate Professor , Department of CSE& IT, Sri Aditya Engineering College, Surampalem, E.G.Dist , Andhra Pradesh, India
asnchakravarthy@yahoo.com
[††]Research Scholar, JNTUK, Kakinada, E.G.Dist , Andhra Pradesh, India
vamsilovesindia@gmail.com
[†††]Professor, Dept. of CS & SE, Andhra University, Visakhapatnam Dist, Andhra Pradesh, India
psavadhani@yahoo.com

## ABSTRACT

*Authentication is the act of confirming the truth of an attribute of a datum or entity. This might involve confirming the identity of a person, tracing the origins of an artefact, ensuring that a product is what it's packaging and labelling claims to be, or assuring that a computer program is a trusted one. The authentication of information can pose special problems (especially man-in-the-middle attacks), and is often wrapped up with authenticating identity. Literary can involve imitating the style of a famous author. If an original manuscript, typewritten text, or recording is available, then the medium itself (or its packaging - anything from a box to e-mail headers) can help prove or disprove the authenticity of the document. The use of digital images of handwritten historical documents has become more popular in recent years. Volunteers around the world now read thousands of these images as part of their indexing process. Handwritten text images of old documents are sometimes difficult to read or noisy due to the preservation of the document and quality of the image [1]. Handwritten text offers challenges that are rarely encountered in machine-printed text. In addition, most problems faced in reading machine-printed text (e.g., character recognition, word segmentation, letter segmentation, etc.) are more severe, in handwritten text. In this paper we Here in this paper we proposed a method for authenticating hand written text images using back propagation algorithm..*

## KEYWORDS
Optical Character Recognition, Handwritten text image, Handwriting Recognition

## 1. INTRODUCTION TO HANDWRITTEN TEXT IMAGE:

Handwriting recognition is the ability of a computer to receive and interpret intelligible handwritten input from sources such as paper documents, photographs, touch-screens and other devices. The image of the written text may be sensed "off line" from a piece of paper by optical scanning (Optical Character Recognition (OCR)) or intelligent word recognition. Alternatively, the movements of the pen tip may be sensed "on line", for example by a pen-based computer screen surface.

### 1.1 Off-Line Hand Writing Recognition:

Off-line handwriting recognition involves the automatic conversion of text in an image into letter codes which are usable within computer and text-processing applications. The data obtained by this form is regarded as a static representation of handwriting. Off-line handwriting recognition is comparatively difficult, as different people have different handwriting styles. And, as of today, OCR engines are primarily focused on machine printed text and ICR for hand "printed" (written in capital letters) text. There is no OCR engine that supports handwriting recognition as of today.

### 1.1.1 Problem domain reduction techniques:

Narrowing the problem domain often helps increase the accuracy of handwriting recognition systems. A form field for a ZIP code for example, would contain only the characters 0-9. This fact would reduce the number of possible identifications.

Primary techniques for problem domain reduction:

> ➢ Specifying specific character ranges

> ➢ Utilization of specialized forms

### 1.1.2 Character extraction:

Off-line character recognition often involves scanning a form or document written sometime in the past. This means the individual characters contained in the scanned image will need to be extracted. Tools exist that are capable of performing this step [1] however, several common imperfections in this step. The most common being characters that are connected together are returned as a single sub-image containing both characters. This causes a major problem in the recognition stage. Yet many algorithms are available that reduce the risk of connected characters.

### 1.1.3 Character recognition:

After the extraction of individual characters occurs recognition engine is used to identify the corresponding computer character. Several different recognition techniques are currently available.

### 1.1.4 Neural networks:

Neural network recognizers learn from an initial image training set. The trained network then makes the character identifications. Each neural network uniquely learns the properties that differentiate training images. It then looks for similar properties in the target image to be identified. Neural networks are quick to setup; however, they can be inaccurate if they learn properties that are not important in the target data.

### 1.1.5 Feature extraction:

Feature extraction works in a similar fashion to neural network recognizers however, programmers must manually determine the properties they feel are important.

Some example properties might be:

> ➢ Aspect Ratio

> ➢ Percent of pixels above horizontal half point

> ➢ Percent of pixels to right of vertical half point

> ➢ Number of strokes

> ➢ Average distance from image centre

> ➢ Is reflected y axis

> ➢ Is reflected x axis

This approach gives the recognizer more control over the properties used in identification. Yet any system using this approach requires substantially more development time than a neural network because the properties are not learned automatically.

## 1.2    Online- Hand Writing Recognition:

On-line handwriting recognition involves the automatic conversion of text as it is written on a special digitizer , where a sensor picks up the pen-tip movements as well as pen-up/pen-down switching. That kind of data is known as digital ink and can be regarded as a dynamic representation of handwriting. The obtained signal is converted into letter codes which are usable within computer and text-processing applications.

The elements of an on-line handwriting recognition interface typically include:

- a pen or stylus for the user to write with.
- a touch sensitive surface, which may be integrated with, or adjacent to, an output display.
- a software application which interprets the movements of the stylus across the writing surface, translating the resulting strokes into digital text.

## 2. CONVERSION OF HAND WRITTEN IMAGE INTO BIPOLAR:

Before giving the image as password the image should be converted in to its Read, Green and Blue (RGB) values and these values will be normalized using our normalization function. As we cannot give the image directly as input to the neural network, here we converted image into matrix (or text).

### 2.1 Conversion of image to matrix (or text):

This method reads the colour of each pixel of the image. Then converts the colour into red, green and blue (RGB) parts as each colour can be produced using these colours.

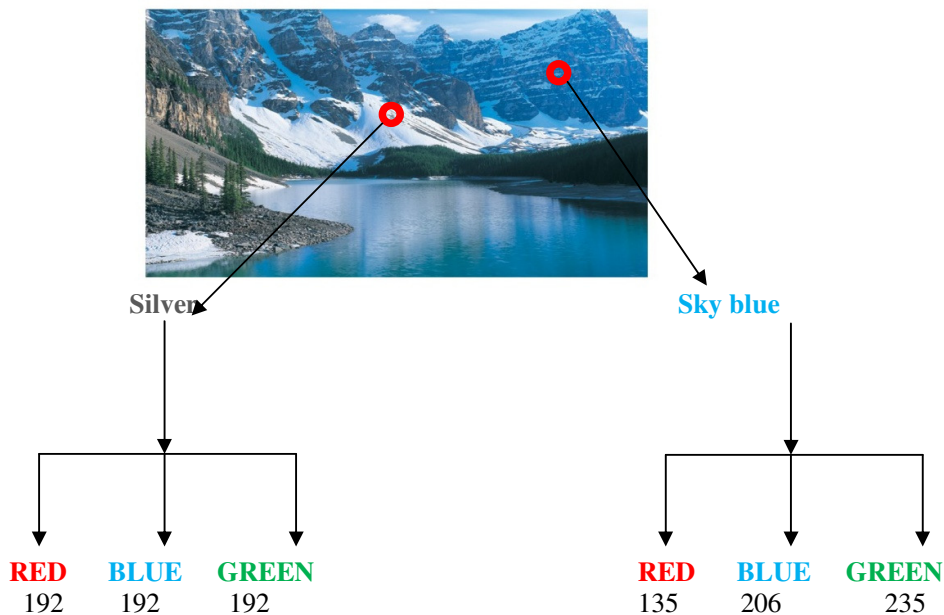

Figure 1. Reading pixel values

By using the above procedure we can convert any image into a matrix consisting of set of numbers representing all the pixels of the image. After converting image into a matrix which consists of set of numbers we can give it to the neural network as input and we can train it using username and image matrix as a training sample. So here user has the choice of selecting either text or an image depending on the requirement and security he expects.

This scheme normalizes numbers in matrix given in figure 3.3 to increase the security. As Red, Blue, Green can have maximum of 255 and minimum of 0, it uses the following formula to normalize the numbers of matrix.

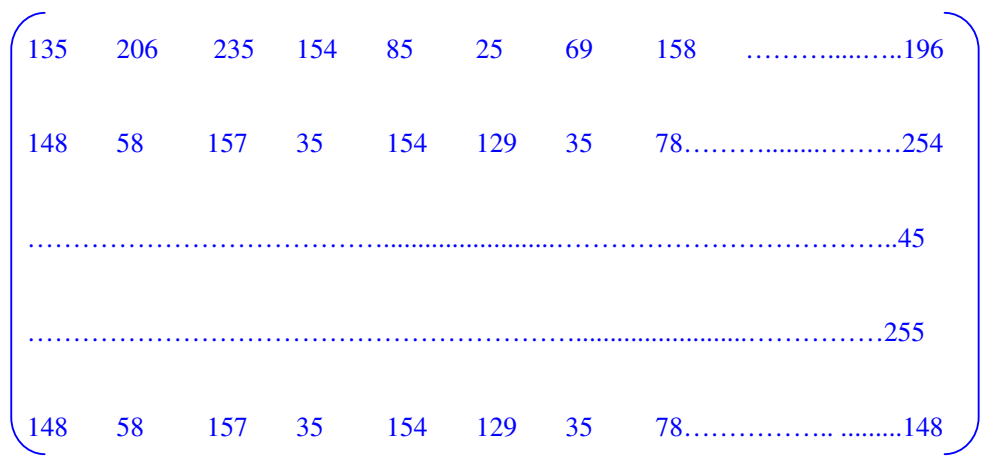

Figure 2. Matrix representation of an image.

$$C_n = \frac{C_t}{255} \qquad\qquad (1)$$

Where $C_t$ is the value of Red or Green or Blue components of the graphical password and255 is the maximum value for any colour.

This method converts the image into text by using procedures discussed above. After converting the image into bipolar values we can use the same procedure which is used for textual password.

## 3. RESULTS FOR HAND WRITTEN IMAGE AUTHENTICATION:

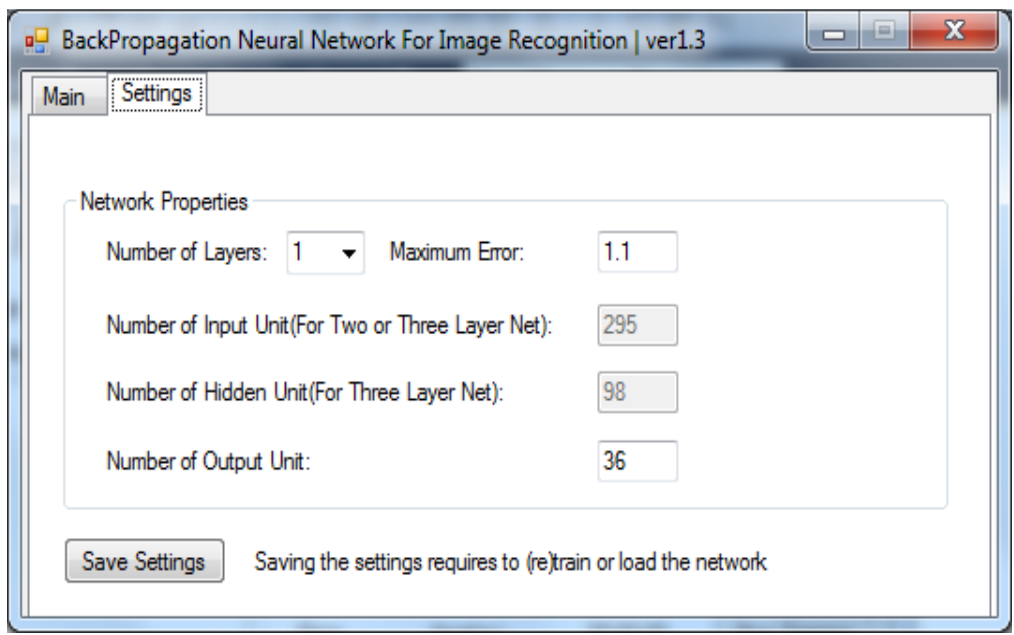

Figure 3. Screen showing how to setup network and back propagation algorithm

In this step *Maximum error* indicates maximum error that we may ignore while training and *No.of patterns* are restricted by no.of output units , In order to recognize more patterns, more number of output units can be used . As shown in figure 7.1 the number layers, maximum error ,number of input units and hidden units will be given and application will save these values for future use.

When number of layers ,number of hidden units and maximum errors are changed the efficiency of password authetication scheme will increase.

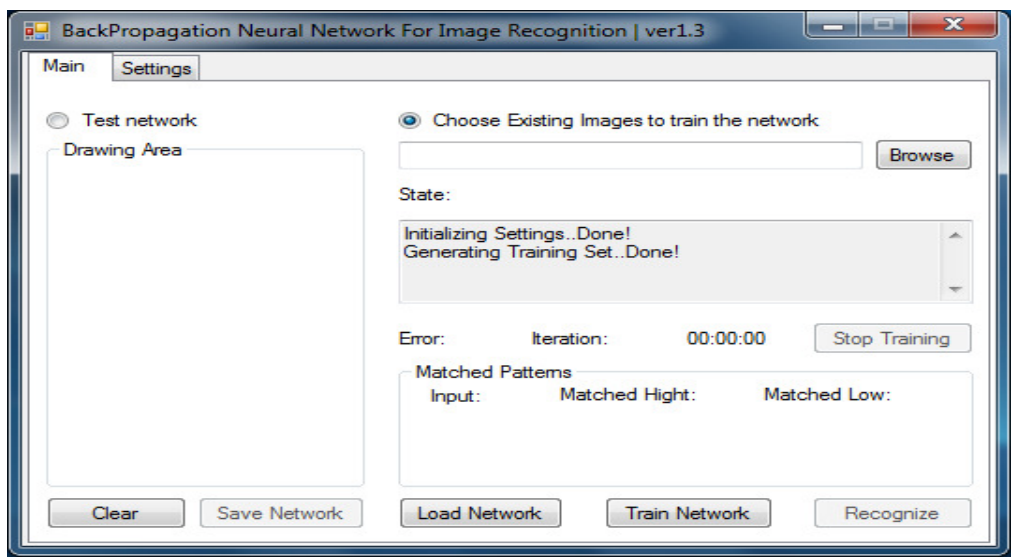

Figure 4. Screen showing specification of the training pattern

The application will provide a facility to choose image for training from a specified folder  as shown in figure 4.
 Once the path is given it will retrieve patterns for training  the network as shown in figure 5.

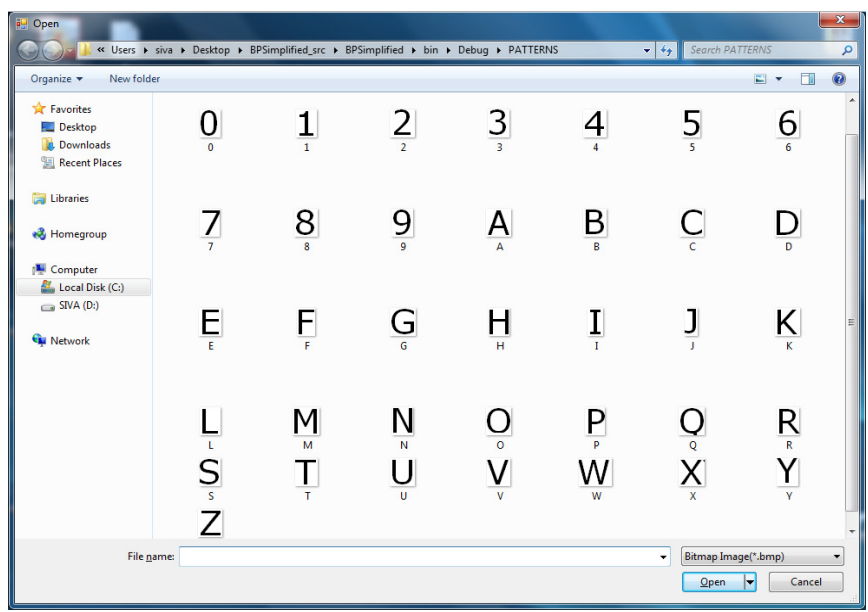

Figure 5. Screen showing how to specify patterns for training the network

## 3.1 Training the BPNN for Hand Written Image:

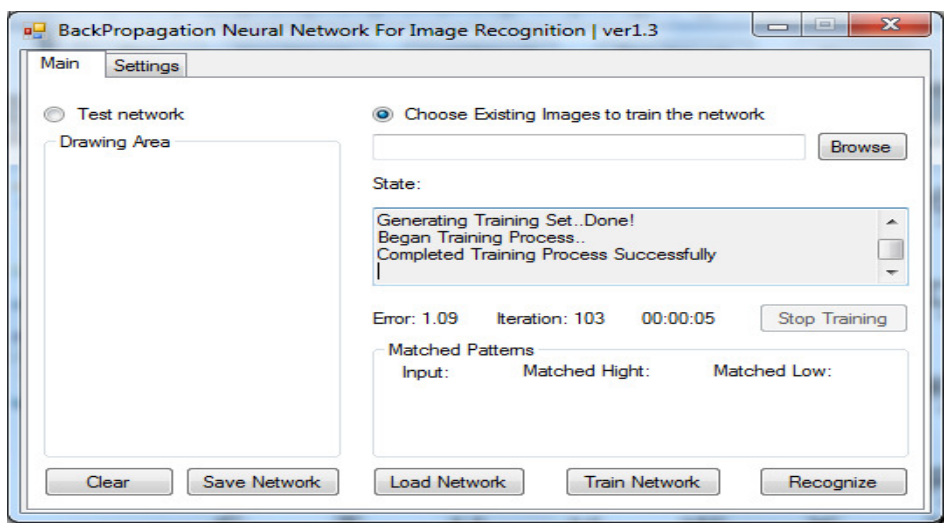

Figure 6. Screen showing how to train the network

In this step as shown in figure 7.4 the error is 1.09 and training completed in 5 seconds.We can decrease error by doing following changes

- By changing the  learning rate paramater
- By changing  the of  input and/or hidden units
- By changing  the number of  layers
- By increasing number of  iterations

## 3.2 Hand written image as input:

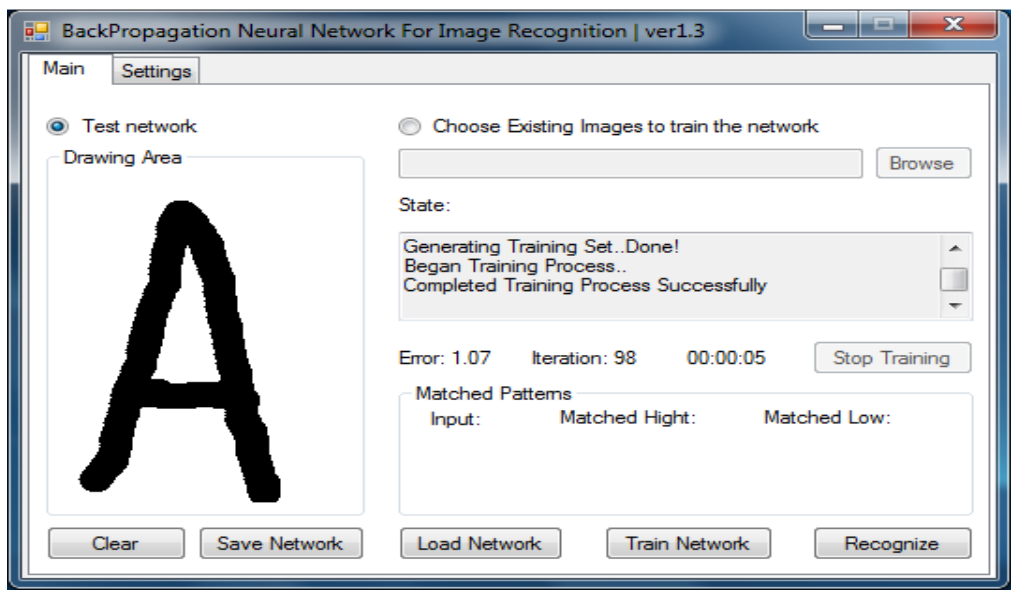

Figure 7. Screen showing how to draw images for recognising

While drawing the image user should take care of width and height of the image. While training the network using two or more similar patterns which represent the same output can avoid restriction on width and height.

## 3.3 Hand Written Pattern recognition:

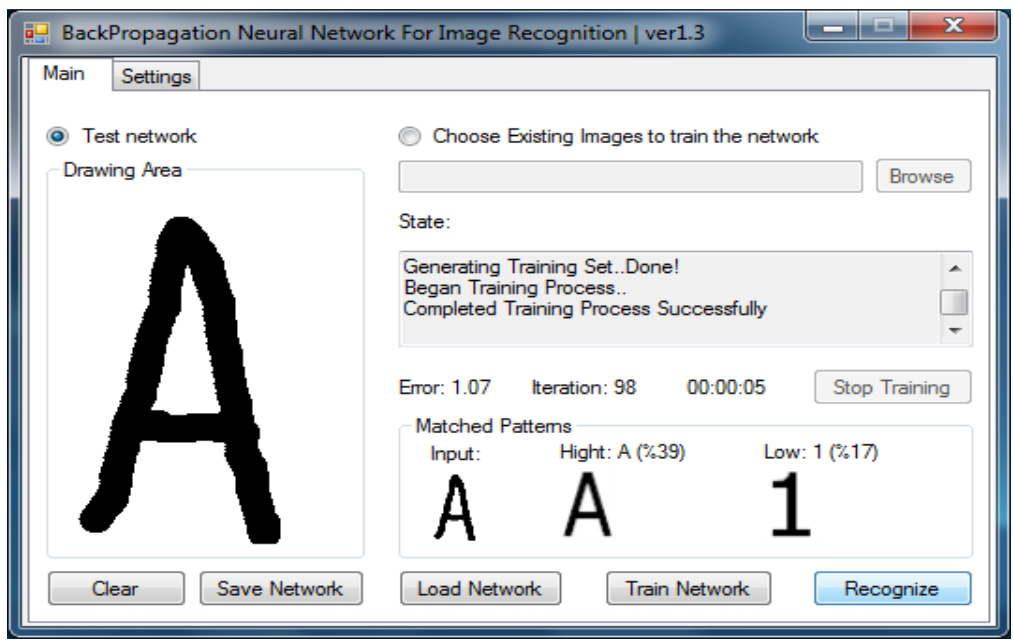

Figure 8. Screen showing how to recognise image

In the figure 8, the input is matched with two patterns, one with low error i.e. **A** and other with high error i.e. **1**. Since **A** is having more similarity(less error) compared to **1**, the application recognizes the given pattern as **A**.

# 4 CONCLUSIONS AND FUTURE SCOPE:

Memory is not just a passive store for holding ideas without changing them; it may transform those ideas when they are being retrieved. There are many examples showing that what is retrieved is different from what was initially stored. A simple model describing context-dependent associative memories generates a good vectorial representation of basic logical calculus. One of the powers of this vectorial representation is the very natural way in which binary matrix operators are capable to compute ambiguous situations. This fact presents a biological interest because of the very natural way in which the human mind is able to take decisions in the presence of uncertainties. Also these memories could be used to develop expert agents to the recent problem domain.

One of the ways in which this technique can be improved is by identifying the start and end of the edges better. Also, a valuable alternative would be to be able to automatically generate the parameters necessary to work with the bilateral filter and the control of noise for the Laplacian filter[1].

This Paper presented a novel password authentication method for the handwritten document images. Unlike the existing approaches using the image-to-image matching-based approaches, we use the word recognition distances to improve the word matching accuracy.

This method estimate the probabilities of word boundary segmentation using the distances between connected components and combine the segmentation and recognition distances to create a probabilistic word matching similarity. This technique shows the improvement obtained by our approach by comparing the image-to-image matching approaches [2, 3] with this method.

The future works can be done to improve the overall performance of PAS by combining multiple systems using different image features and similarity measurements. System combination may also effectively fix the intrinsic drawbacks of every single system.

Next we can consider CAPTCHAs based on handwritten sentence reading and understanding. There are open questions on how long "Handwritten CAPTCHAs" will resist automatic attacks, how robust is our proposed algorithm for image transformation and degradation, or how easily an image deformation can be reversed and the original image retrieved, as well as concerns based on the future technology development of computer vision systems that could eventually fill the gap in ability between humans and machine reading [4].

## ACKNOWLEDGEMENTS

An assemblage of this nature could never have been attempted without reference to and inspiration from the works of others whose details are mentioned in reference section.

## Authors

†A.S.N Chakravarthy received his M.Tech (CSE) from JNTU, Anantapur , Andhra Pradesh, India. Presently he is working as an Associate Professor in Dept. Of Computer Science and Engineering in Sri Aditya Engineering College, SuramPalem, E.G.Dist, AP, India. His research areas include Network Security, Cryptography, Intrusion Detection, Neural networks.

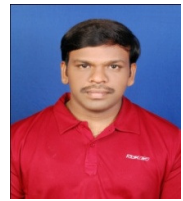

†† Penmetsa V Krishna Raja received his M.Tech (CST) from A.U, Visakhapatnam, Andhra Pradesh, India. He is a research scholar under the supervision of Prof.P.s.Avadhani. His research areas include Network Security, Cryptography, Intrusion Detection, Neural networks.

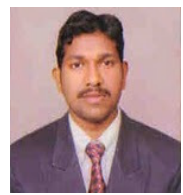

†††Prof. P.S.Avadhani did his Masters Degree and PhD from IIT, Kanpur. He is presently working as Professor in Dept. of Computer Science and Systems Engineering in Andhra University college of Engg., in Visakhapatnam. He has more than 50 papers published in various National / International journals and conferences. His research areas include Cryptography, Data Security, Algorithms, and Computer Graphics, Digital Forensics and Cyber Security.

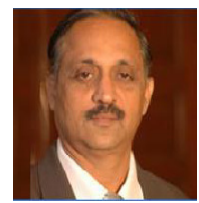

o0o